\documentstyle[twoside,fleqn,npb,epsfig]{article}
\newcommand{\be}{\begin{equation}}
\newcommand{\ee}{\end{equation}}
\newcommand{\bea}{\begin{eqnarray}}
\newcommand{\eea}{\end{eqnarray}}
\newcommand{\beas}{\begin{eqnarray*}}
\newcommand{\eeas}{\end{eqnarray*}}
\newcommand{\bookfig}[4]{
\begin{figure}\centering
\vspace{9pt} \framebox[72mm]{ 
\epsfysize=#4cm \epsfbox{#1}} \caption{ #3}\label{#2}
\end{figure}
}
\newtheorem{theorem}{Theorem}

\newtheorem{prop}[theorem]{Proposition}

\newcommand{\AmS}{{\protect\the\textfont2
  A\kern-.1667em\lower.5ex\hbox{M}\kern-.125emS}}

\title{Unique factorization in perturbative QFT}

\author{D. Kreimer\address{CNRS-IHES\\ 35 rte.\ de Chartres, F-91440 Bures-sur-Yvette, France}%
        \thanks{supported in parts by NSF grant 0205977; Ctr.\ Math.\ Phys.\ at Boston U., BUCMP/02-06}
}

\begin{document}

\begin{abstract}
We discuss factorization of the Dyson--Schwinger equations using
the Lie- and Hopf algebra of graphs. The structure of those
equations allows to introduce a commutative associative product on
1PI graphs. In scalar field theories, this product vanishes if and
only if one of the factors vanishes. Gauge theories are more
subtle: integrality relates to gauge symmetries.
\end{abstract}

\maketitle

\section{The Lie and Hopf algebra of graphs}
Over recent years, the Lie and Hopf algebra structures of Feynman
graphs  have been firmly established in collaboration with Alain
Connes and David Broadhurst
\cite{DK1,Chen,overl,CK1,RHI,RHII,InsElim,BK1,BK,BK2,BK4}.
Hopefully, they find their way now into the algorithmic toolkit of
the practitioner of QFT.  They directly address the computational
practice of momentum space Feynman integrals, but can be
equivalently formulated in coordinate space. They take into
account faithfully the self-similarity of physics at different
scales and the iteration of Green functions in terms of itself by
their quantum equations of motions, the Dyson--Schwinger
equations. We will review quickly these Lie and Hopf algebra
structures and then report in the next section on factorization
properties which exist in the formal series over graphs
contributing to a given Green function. We motivate this
factorization in comparison with the factorization of integers
leading to an Euler product for $\zeta$-functions, and comment on
the fate of such factorizations in gauge theories. Results
reported here can also be found in \cite{annals}, which in
particular contains a more extended discussion of factorization
properties and their implications on perturbative expansions with
regard to the question if the factorization commutes with the
application of the Feynman rules. We start our summary with the
pre-Lie structure.
\subsection{The Pre-Lie Structure}
A 1PI Feynman graph $\Gamma_1$ provides internal edges and
vertices. Each internal vertex $v$ has a set of edges $f_v$
attached to it. A vertex correction graph $\Gamma_2$ has a set of
external edges $\Gamma_{2,{\rm ext}}^{[1]}$ attached to it. If
those two sets agree, we can glue $\Gamma_2$ into $\Gamma_1$ at
$v$, where we sum over all these possible bijections between
$f_{v}$ and $\Gamma_{2,\rm ext}^{1}$, and normalize such that
topologically different graphs are generated precisely once.
Furthermore, we formally define this insertion to vanish when the
two sets do not agree. Summing over all places $v$ in $\Gamma_1$
gives a bilinear map $\ast$ from 1PI graphs to 1PI graphs. It can
be extended to insertions of self-energies into edges in the
obvious manner \cite{RHI,InsElim}. One then has:
\begin{theorem} {\rm \cite{RHI,DK1,InsElim}}
The operation $\ast$ is pre-Lie: \bea [\Gamma_1\ast \Gamma_2]\ast
\Gamma_3   & - &  \Gamma_1\ast[\Gamma_2\ast \Gamma_3]\nonumber\\
=  [\Gamma_1\ast \Gamma_3]\ast \Gamma_2  & - &
\Gamma_1\ast[\Gamma_3\ast \Gamma_2].\eea
\end{theorem}
The theorem says that the lack of associativity in the bilinear
operation $\ast$ is invariant under permutation of the elements
indexed $2,3$. This suffices to show that the antisymmetrization
of this map fulfils a Jacobi identity. Hence we get a Lie algebra
${\cal L}$ by antisymmetrizing this operation: \be
[\Gamma_1,\Gamma_2]=\Gamma_1\ast\Gamma_2-\Gamma_2\ast\Gamma_1,\label{Lie}\ee
and a Hopf algebra ${\cal H}$ as the dual of the universal
enveloping algebra of this Lie algebra \cite{RHI,CK1}. We restrict
attention to graphs which are superficially divergent,  while
superficially convergent graphs can be incorporated in a trivial
manner \cite{RHI}. Fig.(\ref{f2}) gives Lie brackets for various
different theories.  \bookfig{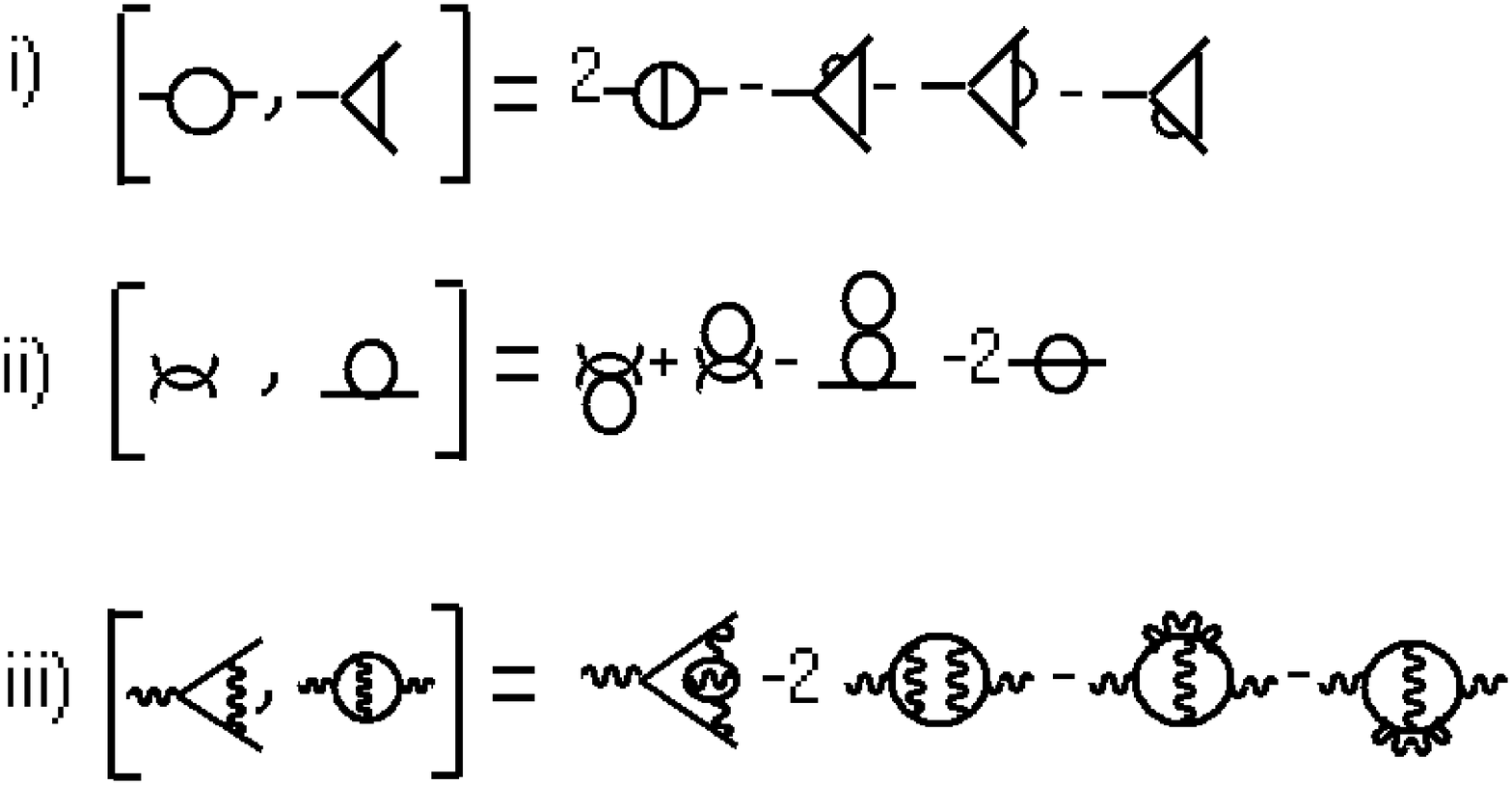}{f2}{Examples of Lie brackets
in various theories: i) $\phi^3_6$ graphs, ii) $\phi^4_4$ graphs,
iii) QED graphs.}{4}

\subsection{The principle of multiplicative subtraction}
The beauty of the existence of these graded combinatorial Lie
algebras is that they make the role of locality manifest. We
dualize the universal enveloping algebra ${\cal U(L)}$ of ${\cal
L}$ and obtain a commutative,  non-cocommutative Hopf algebra
${\cal H}$ \cite{RHI}. To find this dual, one uses a Kronecker
pairing and constructs it in accordance with the Milnor-Moore
theorem \cite{RHI,InsElim,CK1}.

1PI graphs are the linear generators of the Hopf algebra, with
their disjoint union furnishing a commutative product in the
algebra. We then identify the Hopf algebra by a coproduct
$\Delta:{\cal H}\to {\cal H}\otimes{\cal H}$:
\be\Delta(\Gamma)=\Gamma\otimes
1+1\otimes\Gamma+\sum_{\gamma{\subset}
\Gamma}\gamma\otimes\Gamma/\gamma,\ee the sum is over all unions
of one-particle irreducible (1PI) superficially divergent proper
subgraphs. We extend this definition to products of graphs so that
we get a bialgebra. The above sum should, when needed, also run
over appropriate projections to formfactors, to specify the
appropriate type of local insertion \cite{RHI} which appear in
local counterterms, which we omitted in the above sum for
simplicity. Fig.(\ref{f3}) gives  examples.
\bookfig{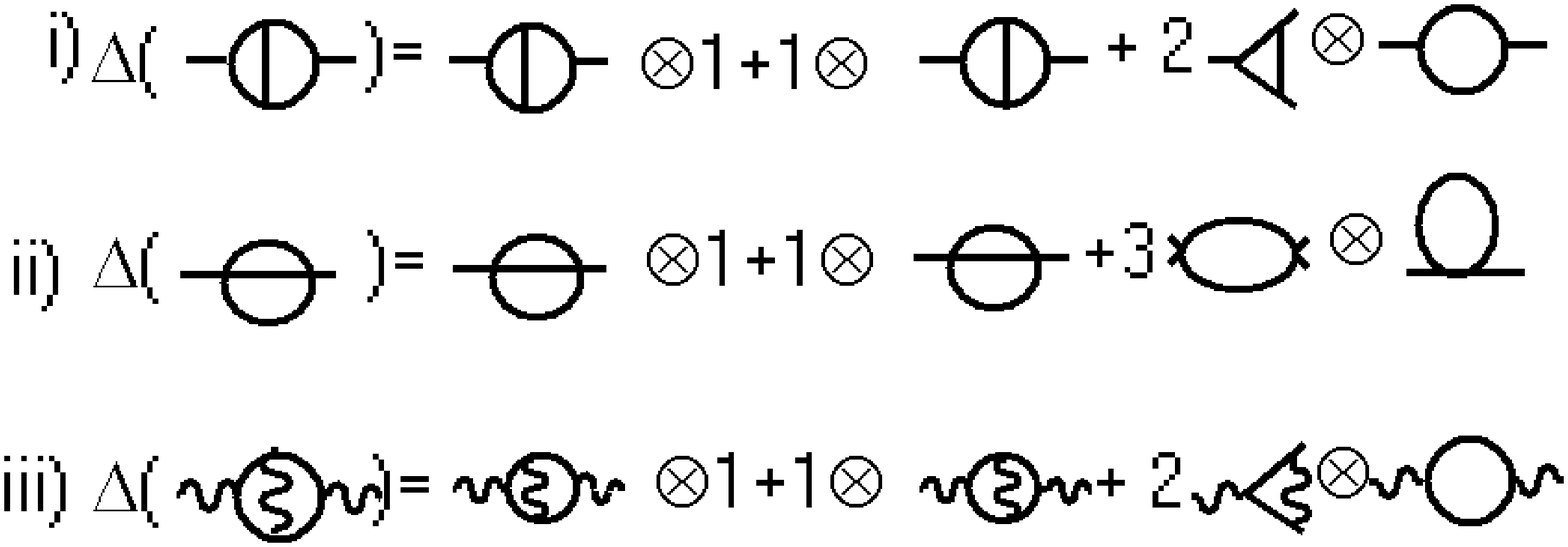}{f3}{Examples for  coproducts $\Delta(\Gamma)$ in
various theories: i) $\phi^3_6$, ii) $\phi^4_4$, iii) QED.}{2.8}

We still need a counit and antipode (coinverse): the counit
$\bar{e}$ vanishes on any non-trivial Hopf algebra element,
$\bar{e}(1)=1,\,\bar{e}(X)=0$. At this stage we have a
commutative, but typically not cocommutative bialgebra. It
actually is a Hopf algebra as the antipode (coinverse) in such
circumstances comes for free as \be
S(\Gamma)=-\Gamma-\sum_{\gamma{\subset}
\Gamma}S(\gamma)\Gamma/\gamma.\ee Those ingredients give us a
group structure on any multiplicative map from the Hopf algebra to
some ring or algebra $V$: the co-unit, -product, -inverse give us
a unit, product and inverse on such maps.

As physicists, we have a distinguished such map, (momentum-space)
Feynman rules $\phi:{\cal H}\to V$ from the Hopf algebra of graphs
${\cal H}$ into an appropriate space $V$, of, say, Laurent
polynomials in the dimensional regularization parameter
$\varepsilon$.

We will have to make one further choice: a renormalization scheme.
For us, this is a map $R:V\to V$ faithful on short distance
singularities, which furthermore obeys \be
R(xy)+R(x)R(y)=R(R(x)y)+R(xR(y)),\label{RB}\ee
 an equation which guarantees the
multiplicativity of renormalization and is at the heart of the
Birkhoff decomposition of \cite{RHI,RHII}.

Renormalization theory emerges as a principle of multiplicative
subtraction: we define a further character $S_R^\phi$ which
deforms $\phi\circ S$ slightly and delivers the counterterm for
$\Gamma$ in the renormalization scheme $R$: \be
S_R^\phi(\Gamma)=-R[\phi(\Gamma)]-R\left[\sum_{\gamma{\subset}
\Gamma}S_R^\phi(\gamma)\phi(\Gamma/\gamma)\right]\ee which should
be compared with the undeformed \be \phi\circ
S=-\phi(\Gamma)-\sum_{\gamma{\subset} \Gamma}\phi\circ S
(\gamma)\phi(\Gamma/\gamma).\ee  We obtain the renormalization of
$\Gamma$ by the application of a renormalized character
$$\Gamma\to S_R^\phi\star\phi(\Gamma)$$ and Bogoliubov's $\bar{R}$
operation as \be\bar{R}(\Gamma)=\phi(\Gamma)+
\sum_{\gamma{\subset}\Gamma}
S_R^\phi(\gamma)\phi(\Gamma/\gamma),\ee so that we have \be
S_R^\phi\star\phi(\Gamma)=\bar{R}(\Gamma)+S_R^\phi(\Gamma).\ee
Here, $S_R^\phi\star\phi$ is an element in the group of characters
of the Hopf algebra, with the previously announced group law given
by
$$\phi_1\star\phi_2=m_V\circ(\phi_1\otimes\phi_2)\circ\Delta,$$
so that indeed the coproduct, counit and coinverse (the antipode)
give the product, unit and inverse of this group, as it befits a
Hopf algebra. This Lie group has the previous Lie algebra ${\cal
L}$ of graph insertions as its Lie algebra \cite{RHI}.

\section{Factorization}
There is a connection between Dy\-son--Schwin\-ger equations and
Euler products \cite{annals}. To motivate it let us start with the
Riemann $\zeta$ function, and obtain it as a solution to a
Dyson--Schwinger equation, as in \cite{annals}. This is only meant
as a sufficient stimulus to invert the reasoning and look for
Euler products in quantum field theory. This whole section is
meant to inspire an investigation of the question how such
factorizations of Dyson--Schwinger equations actually fare when we
apply Feynman rules or related characters to Euler products, a
question addressing the transition from the perturbative to the
non-perturbative \cite{annals}. It will be discussed in detail in
upcoming work of the author.
\subsection{The Riemann $\zeta$-function from a Dy\-son-Schwin\-ger
equation} The Riemann $\zeta$-function is the analytic
continuation of the sum $\sum_n 1/n^s$, and has an Euler product
\be \zeta(s)=\sum_{n}\frac{1}{n^s}=\prod_p\frac{1}{1-p^{-s}},\;
\Re (s)
>1,
\ee over all primes $p$ of the (rational) integers.

There is a well-known  Hopf algebra of sequences
$(p_1,\ldots,p_k)$ (the $p_i$ are primes), and we can introduce
$B_+^p[J]$ as the sequence which is obtained by adding a new prime
$p$ as the first element to the sequence $J$. The Hopf algebra
structure emerges when we require that $B_+^p$ is Hochschild
closed for all $p$ \cite{annals,CK1}: \be
\Delta(B_+^p[J])=B_+^p[J]\otimes 1+[{\rm id}\otimes
B_+^p]\Delta[J],\ee with $\Delta(1)=1\otimes 1$ and we identify
$1$ with the empty sequence. Define the value $w(J)$ to be the
product of the entries of $J$, and let the symmetry factor $S(J)$
be the number of sequences which have the same value, which simply
is $k!$ if the sequence has length $l(J)=k$.  Note that for a one
element sequence $(p)$, \be \Delta[(p)]=(p)\otimes 1+1\otimes
(p),\ee primitive elements have prime value, $w((p))=p$.

Consider the "Dyson--Schwinger equation" \be
\overline{\zeta}(\rho)=1+\rho\sum_p
B_+^p[\overline{\zeta}(\rho)].\ee We obtain a formal series (in
"the coupling" $\rho$) \be \overline{\zeta}(\rho)=1+\rho\sum_p
(p)+\rho^2\sum_{p_1,p_2}(p_1,p_2)+\cdots . \ee Define "Feynman
rules" by $\phi_s(J)=\frac{1}{l(J)!}w(J)^{-s}$, and set \be
\zeta(s,\rho)=\phi_s[\overline{\zeta}(\rho)]. \ee We regain
Riemann's $\zeta$ function as \be \zeta(s)=\lim_{\rho\to
1}\zeta(s,\rho). \ee

Note the general structure of the formal "Dy\-son--Schwin\-ger
equation" above: it determines an unknown $\overline{\zeta}(\rho)$
in terms of itself, as "1 plus a sum over the image of the unknown
$\overline{\zeta}(\rho)$ under all closed Hochschild one cocycles
$B_+^p$, weighted by appropriate symmetry factors". Is there an
Euler product for $\overline{\zeta}$?

The simplest way is to get it from the well-known shuffle product
on sequences. \bea B_+^{p_1}(J_1)\sqcup B_+^{p_2}(J_2) & =&
B_+^{p_1}(J_1\sqcup
B_+^{p_2}(J_2))\nonumber\\
& + & B_+^{p_2}(B_+^{p_1}(J_1)\sqcup J_2). \eea Then, \be
\overline{\zeta}(\rho)=\Pi_p^\sqcup\frac{1}{1-\rho\; (p)},\ee
 $(1-\rho\;(p))^{-1}=1+\rho\;(p)+\rho^2\;
(p)\sqcup(p)+\cdots,$ and where the shuffle product is used in the
Euler product throughout. We have \be\zeta(s)=\phi_s\left(
\Pi_p^\sqcup\frac{1}{1-\rho\; (p)}\right) =\Pi_p
\frac{1}{1-p^{-s}},\ee the evaluation of the product is the
product of the evaluations.

The reason we dared calling the above equation a Dyson--Schwinger
equation is a simple fact - the true Dyson--Schwinger equations of
QFT have a similar structure: they express an unknown Green
function as a sum over all possible insertions of itself in all
possible skeleton diagrams. This allows to write the unknown Green
function as a sum over all possible images over all closed
Hochschild one-cocycles in the theory \cite{annals,dennis},
precisely provided by the primitive bidegree one graphs $\gamma$,
which play the role of primes.

\subsection{Factorizing graphs}
Can a factorization into Euler products  be found in quantum field
theory? And if it can be found at the combinatorial level, will
the evaluation, by the Feynman rules, equal the product of the
evaluations, and, if not, how  will it deviate?

A typical Dyson--Schwinger equation is of the form \cite{annals}
\be X=1+\sum_\gamma B_+^\gamma(g^k  [\cup_k X]),\ee where the
infinite sum in the Hopf algebra is over primitive graphs
$\gamma$, $k=k(\gamma)$ is the loop-degree of $\gamma$, and as the
notation indicates, the maps $B_+^\gamma$ are closed Hochschild
one-cocycles, and the sum is over all of those \cite{annals,CK1}.
$X$ is an infinite sum of graphs contributing to a chosen Green
function, and evaluation by the Feunman rules delivers the usual
Dyson--Schwinger equations given as an integral equation over the
kernels provided by the primitive graphs $\gamma$. Note that, as
insertion into a primitive graph commutes with the coproduct in
the desired way, we can directly read-off the renormalized
Dyson--Schwinger equation as \be X_R=Z_X+\sum_\gamma
B_+^\gamma(g^k  [\cup_k X_R]),\ee where the $Z$-factor $Z_X$ is
the negative part in the Birkhoff decomposition with respect to a
renormalization scheme $R$, as usual.

The first question to ask is: is there a combinatorial  Euler
product for the formal sum over graphs generated by such an
equation? The answer is indeed affirmative.

 The crucial step lies in the definition of an associative
product $\vee$ which generalizes the shuffle product $\sqcup$
appropriate for totally ordered sequences to the partial order
given by being a subgraph \cite{annals}. We simply describe it in
Fig.(\ref{f4}). \bookfig{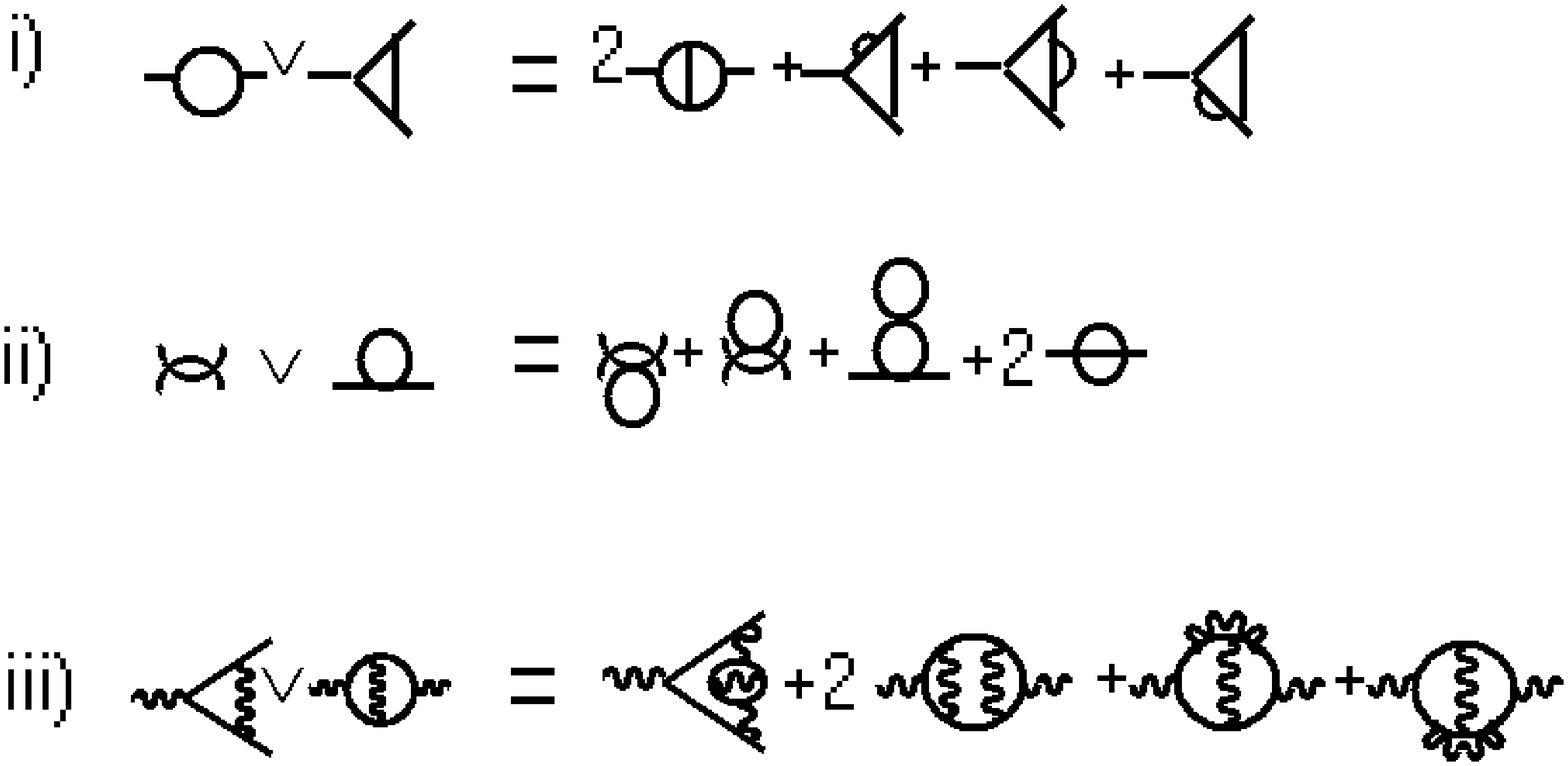}{f4}{Examples of $\vee$-products
in various theories: i) $\phi^3_6$, ii) $\phi^4_4$, iii) QED. In
these simple examples, they correspond just to a symmetrization
instead of an antisymmetrization of the pre-Lie insertion. The
general definition is given in \cite{annals}.}{3.7}

Then, we have
\begin{theorem}$X(g)=\prod^\vee_\gamma \frac{1}{1-g^{k(\gamma)}\gamma},$\end{theorem}
the proof of which is to be given elsewhere.

Here is not the space to discuss the fascinating question of how
much we can  say about
$$ \phi\left(\prod^\vee_\gamma
\frac{1}{1-\gamma}\right)\; \mbox{vs}\; \prod_\gamma
\frac{1}{1-\phi(\gamma)}=\zeta_G(\phi)\;?$$ A discussion of this
question was started in \cite{annals}, and answers are in
preparation. Note that such factorization suggest
pseudo-exponentiations of transcendentals in the Taylor expansion
of skeleton graphs in the dimensional regularization parameter, as
reported by David Broadhurst at this conference \cite{pseudo}.

\subsection{Gauge symmetries}
Finally, let us mention a first simple example as to how basic
algebraic structures relating to such factorizations in
Dyson--Schwinger equations relate to physical properties of a
theory. Let $\gamma_{\rm vp}$ denote the one-loop
vacuum-polarization in QED, and more generally let $\{ \Gamma_{\rm
N-A}\}$ be the set of 1PI graphs in a non-abelian gauge theory
which have as external lines two or more gauge bosons and such
that the three-valent fermion-boson vertex is the only type of
vertex which appears in them. They hence all have at least one
fermion loop.

\begin{prop}
i) The product $\Gamma_1\vee\Gamma_2$ is integral for 1PI graphs
in $\phi^3_6$
and $\phi^4_4$.\\
ii) It is non-integral for gauge theories:\\
$\Gamma_1\vee\Gamma_2=0\Rightarrow $ $ \Gamma_1=0$ or $\Gamma_2=0$
or $\Gamma_1=\Gamma_2=\gamma_{\rm vp}$ (QED);\\
$\Gamma_1\vee\Gamma_2=0\Rightarrow $ $ \Gamma_1=0$ or $\Gamma_2=0$
or $\Gamma_1,\Gamma_2\in \{\Gamma_{\rm N-A}\}$ (non-abelian
case).\end{prop} Proof: i) Each 1PI graph with one or more loops
contains internal edges and vertices. As these theories are
renormalizable and we restrict the Hopf algebra to superficially
divergent graphs, the product $\vee$, symmetrizing over insertions
of superficially divergent graphs into each other, will not
vanish. ii) QED: the one-loop vacuum polarization has two internal
fermion propagators and two internal vertices, but no internal
photon line. But it is itself a self-energy graph for the photon.
Its $\vee$-product with itself hence vanishes. Each other QED
graph $\Gamma$ has internal vertices, fermion lines and photon
lines, so that $\Gamma\vee\gamma=0$ implies $\gamma=0$. For the
non-abelian case, this argument has a natural generalization to
the indicated set of graphs, and relates to questions of
uniqueness of factorization \cite{dennis}.

This relates integrality of the $\vee$-product to  gauge/BRST
symmetries. For example, the Hopf algebra of QED graphs can be
divided by an appropriate ideal of graphs such that the quotient
consists of graphs free of one-loop vacuumpolarization subgraphs,
and this quotient turns out to be equivalent to the ideal
generated by the Ward identities. This deserves a much more
detailed discussion again to be given elsewhere.

\end{document}